# METHOD OF THE MULTIDIMENSIONAL SIEVE IN THE PRACTICAL REALIZATION OF SOME COMBINATORIAL ALGORITHMS

Krasimir Yankov Yordzhev
South-West University
Blagoevgrad

Ana Georgieva Markovska
South-West University
Blagoevgrad

**Abstract**    *Some difficulties regarding the application of the well-known sieve method are considered in the case when a practical (program) realization of selecting elements, having a particular property among the elements of a set with a sufficiently great cardinal number(cardinality). In this paper the problem has been resolved by using a modified version of the method, utilizing multidimensional arrays. As a theoretical illustration of the method of the multidimensional sieve, the problem of obtaining a single representative of each equivalence class with respect to a given relation of equivalence and obtaining the cardinality of the respective factor set is considered with relevant mathematical proofs.*

## INTRODUCTION

Let us consider the following class of problems, often met in informatics: A finite set M is given with a family of its subsets $M_\alpha \subset M, \alpha \in I, |I| = N < \infty$, such that $\bigcup_{\alpha \in I} M_\alpha = M$, knowing that for each $\alpha \in I$ an element $x \in M_\alpha$ exists, for which $x \notin M_\beta$ is valid when $\beta \neq \alpha$. Such an element will be referred to as a *"characteristic representative"* of subset $M_\alpha \subset M$. We also suppose that for each $\alpha \in I$ the set $M_\alpha$ is unique for each of its characteristic representatives $x$ and that it can be completely determined by $x$. In other words, we suppose that an effective algorithm exists, which can obtain $M_\alpha$, when an arbitrary characteristic representative is given at the input.

For example, let *M* be the closed interval *[2,s]* where *s* is a natural number, and let *I* be the subset of all the prime numbers $\alpha \in [2, s]$; let us denote by $M_\alpha$ the set of all numbers from *M* which can be divided by $\alpha$. Then for each $\alpha \in I$ there exists a unique characteristic representative of the set $M_\alpha$ and it is simply the prime number $\alpha$. It is possible that $M_\alpha$ may contain more than one characteristic representative. For example, if $\sigma$ is a non-trivial (namely different from the relation equation) relation of equivalence in *M* and $M_\alpha$ are the equivalence classes with respect to $\sigma$, then each element of *M* is a characteristic representative of a certain equivalence class.

A typical problem in computer programming is to obtain (at least) one set $\overline{M} \subseteq M$ which contains a single characteristic representative of each subset of $M_\alpha$, $\alpha \in I$. As a consequence of this problem follows the combinatorial problem of finding the cardinality $N=|I|$ of the index set I. In case that $\sigma$ is a relation of equivalence, then the given problem solves the problem of the cardinality of the factor set $M/_\sigma$.

*Boolean (or Binary, or (0,1)-matrix)* is a matrix whose elements are equal to zero or one. Let $B_{m \times n}$ be the set of all $m \times n$ boolean matrices. It is well known that

(1) $$|B_{m \times n}| = 2^{mn}$$

Let $X, Y \in B_{m \times n}$. An equivalence relation $\rho$ is defined as follows: $X\rho Y$ if and only if X can be obtained from Y by a sequential moving of the last row or column to the first place.

The goal of this paper is to describe an effective algorithm for finding the number of elements of the factor set $\tilde{B}_{m \times n} = B_{m \times n/\rho}$, as well as finding a characteristic representative of each equivalence class. Here we will describe an algorithm which is a modification of the well-known method, known as the "Sieve of Eratosthenes", and which overcomes some difficulties which would inevitably arise with sufficiently great $m$ and $n$ if we apply the classical version. The main difficulty to be overcome arises from the great number of elements of $B_{m \times n}$ with comparatively small $m$ and $n$, according to (1).

In [9] an algorithm is shown, which utilizes theoretical graphical methods for finding the factor set $\tilde{S}_n = S_{n/\rho}$, where $S_n \in B_{n \times n}$ is a set of all permutation matrices, i.e. Boolean matrices having exactly one 1 on each row and each column.

The equivalence classes of $B_{m \times n}$ by the equivalence relation $\rho$ are called double coset ( see [4] § 1.7 or [6] v. 1,ch 2 §1.1). They make use of substitution groups theory ( see[5] § 1.12, § 2.6) and linear representation of finite groups theory ( see [3] § 44-45).

The elements of the set $\tilde{B}_{m \times n} = B_{m \times n/\rho}$ put carry into practice in the textile technology [2].

For undefined notions and definitions we refer to [7], [1] or [8].

**STATEMENT**

**1. Method of the Sieve**

Let a random characteristic element $x$ be given for the set $M_\alpha \subset M$, $\alpha \in I$ and let there also be given effective algorithms estimating the functions
(2) $\text{Next}_1(x)$, $\text{Next}_2(x)$, … $\text{Next}_k(x)$,
defined in $M$ and with the help of which we can obtain the whole set $M_\alpha$. By definition for each $i=1,2,...,k$ we put:
(3) $\text{Next}_i^0(x) = x$
(4) $\quad\quad\quad \text{Next}_i^1(x) = \text{Next}_i(x)$
(5) $\quad\quad\quad \text{Next}_i^t(x) = \text{Next}_i(\text{Next}_i^{t-1}(x))$

when $t \geq 2$, t – integer
Since $M$ is a finite set, then it follows that for each $i=1,2,...,k$ and for each $x \in M$ there exists a minimum natural number
(6) $\quad\quad\quad r(i,x)$
such that for each positive integer number $z$
$\text{Next}_i^{r(i,x)+z} \in \bigcup_{t=0}^{r(i,x)} \text{Next}_i^t(x)$ or $\text{Next}_i^{r(i,x)+z}(x)$ falls out of the range of $M$, or $\text{Next}_i^{r(i,x)+z}(x)$ is not defined in $M$.
If $x$ is a characteristic representative of the set $M_\alpha \subset M$, then we put

(7) $\quad\quad\quad M_1(x) = \bigcup_{t=0}^{r(1,x)} \text{Next}_1^t(x)$

(8) $\quad\quad\quad M_i(x) = \bigcup_{y \in M_{i-1}} \left[ \bigcup_{t=0}^{r(i,y)} \text{Next}_i^t(y) \right]$

for i= 2,3,…,k
Since the union in the square brackets of (8) begins at t=0 and according to (3) it is easily seen that
(9) $\quad\quad M_1(x) \subseteq M_2(x) \subseteq ... \subseteq M_k(x)$

We suppose that the functions $\text{Next}_i(x)$, $i=1,2,...,k$ are chosen so that going from the characteristic element $x \in M_\alpha$ and using the formulae (3) ÷ (9) we obtain the set $M_\alpha$, and we have
(10) $\quad\quad M_\alpha = M_k$
The known method of the sieve we can describe using the following summarized algorithm:
**Algorithm 1** Method of the Sieve
*1. In M we introduce an order (this is always possible, since M is finite) and we sort it according to this order. Let us denote by c#(x) the consecutive number of $x \in M$ according to this order.*

*2. We declare a (one-dimensional) Boolean array H with $m = |M|$ elements. The elements of H will be indexed by the elements of M, i.e. with H[x] we shall denote the element from the array H which corresponds to x from M. (In practice this means that we have numbered the elements from H using the function #(x).)*

*3. Initially we take all elements of H to be zero. Later on, in case we change an H[x] to 1, then this will mean that we have "crossed out" an x.*

*4. We declare the counter N, which is initialized by 0. In case of normal termination of the algorithm, N will be showing the cardinality of the index set I.*

*5. We declare the variable w and we take it to be zero. Variable w will "remember" the consecutive number of the last found*

*characteristic element of the respective subset $M_\alpha$ (The algorithm will discover just one characteristic element for each subset $M_\alpha \subset M$.)*

*6. If an element x such that #(x)>N and H[x]=0 does not exist, then the algorithm terminates. All elements of H for which H[x]=0 will correspond to the elements of the set $\overline{M}$, which contains just one characteristic representative of each subset $M_\alpha \subset M$, $\bigcup_{\alpha \in I} M_\alpha = M$ and which contains only such elements. N will be equal to the cardinality of $\overline{M}$ ( from which it follows that it will also be equal to the cardinality of the index set I), i.e. to the number of non-zero elements of the array H.*
*else*
*7. We find the minimum $x \in M$, such that #(x)>w and H[x]=0.*
*8. We take w=#(x).*
*9. We increase N by one.*
*10. We obtain the set $M_1$ according to the formula (7).*
*11. For each i=2,3,...,k we obtain the sets $M_i$ according to the formula (8). According to (10) we have obtained $M_\alpha = M_k$*
*12. For each $y \in M_\alpha \setminus \{x\}$ we assign H[y] = 1*
*13. We return to point 6.*

In particular, if in Algorithm 1 the set *M* is an ordered set of the natural numbers in the interval *[2,s]*, by putting *k=1* and defining recursively the function
$Next_1^0(x) = x$
and $Next_1^t = Next_1^{t-1}(x) + x$ for *t=1,2,3,...*,
then we obtain the well-known ancient algorithm for finding all prime numbers in the interval *[2,s]*, known by the name the "Sieve of Eratosthenes".

A number of applications of the method of the sieve for solving various problems is described in [7].

**2. The Method of the Multidimensional Sieve**

The method described in part 1 has a number of disadvantages, the main of which is that it is practically inapplicable for programs when a sufficiently great number of elements is present in the base set *M*. This limitation comes from the maximum integer(a number written with a fixed decimal point) which can be used in the corresponding programming environment. For example, by standard in the C++ language the biggest number of the type unsigned long int is equal to $2^{32} - 1$, which in a number of cases is insufficient for the previously defined array *H* to be completely addressed. For example, if the base set $M = B_{m \times n}$, then from (1) it follows that for relatively small m and n the previously described method is impossible to be realized in this language without "special tricks". If we choose, for example, between all 6 x 6 Boolean matrices those which have a given property, for the classical sieve method it is necessary to declare a one-dimentional array with dimensions $2^{36}$, which is significantly greater than the maximum integer number which can be used as an address of an array with the widely distributed translators and programming environments. Here this will be avoided by using a multidimensional Boolean array, the elements of which have a one-to-one correspondence to the elements of the base set, with a much smaller range of the indices.

Let us denote by $Z_{u,v}$, where *u* and *v* are natural numbers and $u \leq v$, the set $Z_{u,v} = \{u, u+1, ... v\}$. The essence of the method, which we refer to as the multidimensional sieve method, is to find such numbers $u_1, u_2,..., u_m, v_1, v_2, . v_m$, for which a one-to-one correspondence between the base set *M* and the Cartesian product $Z_{u_1,v_1} \times Z_{u_2,v_2} \times ... \times Z_{u_m,v_m}$ exists. Then in Algorithm 1 instead of the one-dimensional Boolean array *H* we will declare and then go around and work with an *m*-dimensional Boolean array *W*, the *i*-th index ($1 \leq i \leq m$) of which will vary from $u_i$ to $v_i$. In this way we will reduce the number with which indexing will be done. In the example which we consider we prove that this reduction can be significant. With the elements of the array *W* we will encode the elements of *M*, according to the previously mentioned one-to-one correspondence. For the ordering in *W* (as in this way according to the one-to-one correspondence we introduce ordering in *M*) it seems that the most natural orderings theto be a lexicographic ordering.

As an illustrative example we will consider the following
**Problem 1** *Let $B_{m \times n}$ be the set of all $m \times n$ Boolean matrices and let A, $B \in B_{m \times n}$. We will state that $A \rho B$ if B can be obtained from A by moving the last row or column to the first place, if the other rows are moved one row below and the other columns - one column to the right.*

*Prove that $\rho$ is a relation of equivalence. Find the cardinality $|B_{m \times n / \rho}|$ of the factor set $B_{m \times n / \rho}$ and show one representative of each equivalence class.*

The proof that $\rho$ is a relation of equivalence is trivial and we will omit it here. The authors of this paper are not familiar with an existing common formula for finding $|B_{m \times n / \rho}|$. Here we have set ourselves the simpler task to describe an algorithm for finding $|B_{m \times n / \rho}|$ when specific m and n are given, and show one representative of each equivalence class. The algorithm is based on the notes in the beginning of this section, and on the following two theorems.

**Theorem 1** *Let us denote by $P_n$ the set*

(11) $\quad P_n = \{0, 1, ..., 2^n - 1\}$

*Then a one-to-one correspondence (bijection) between the elements of the Cartesian product $P_n^m = P_n \times P_n \times ... \times P_n$ and the elements of the set $B_{m \times n}$ of all $m \times n$ Boolean matrices exists.*

**Proof**. We consider the image $\alpha : P_n^m \to B_{m \times n}$, defined in the following way: If $\pi = <p_1, p_2, ..., p_m> \in P_n^m$,
then let us denote by $z_i$, $i=1,2,...,m$, the representation of the number $p_i$ in a binary system, and if less than $n$ digits(0 or 1) are necessary, we fill from the left with insignificant zeros, so that $z_i$ will be written with exactly $n$ digits. Since by definition $p_i \in P_n$, i.e. $0 \leq p_i \leq 2^n - 1$, this will always be possible. Then we form an $m \times n$ Boolean matrix, so that the $i$-th row is $z_i$, $i=1,2,...m$. Apparently this is a correctly defined image of $P_n^m$ in $B_{m \times n}$ It is clear that for different $\pi$-s from $P_n^m$ with the help of $\alpha$ we will obtain different matrices from $B_{m \times n}$, i.e. $\alpha$ is injection. Conversely, rows of each Boolean matrix can be considered as natural numbers, written in binary system by using exactly $n$ digits 0 or 1, eventually with insignificant zeros in the beginning, that is, these numbers belong to the set $P_n = \{0, 1, ..., 2^n - 1\}$. Therefore each $m \times n$ Boolean matrix corresponds to an m-tuple of numbers $<p_1, p_2, ..., p_m> \in P_m^n$, that is,
$\alpha$ is surjection. Hence $\alpha$ is a one-to-one correspondence(bijection).

It is easy to see the validity of the following statement, which in fact shows the meaning of our considerations.

**Proposition 1** *Let us denote by $\mu$ the maximum number which is used when coding the elements of the set $B_{m \times n}$ by means of the bijection, defined in theorem 1. Then, for sufficiently great m and n, the following is valid:*
(12) $\quad \mu = \max(2^n - 1, m) << |B_{m \times n}| = 2^{mn}$

Let *a* and *b* be natural numbers. With *a/b* we will denote the operation "integer number division" of *a* and *b*, i.e. if the division has a remainder, then the fractional part is cut, and with *a%b* we will denote the remainder when dividing *a* by *b*.

We consider the function
(13) $\quad \xi(a) = (a\%2)2^{n-1} + a/2$
where % and / are the defined in the previous paragraph operations with integer numbers.

**Theorem 2** *Let $\alpha$ be the defined in the proof of Theorem 1 bijection and let the functions $f_r, f_c : P_n^m \to P_n^m$ be defined in the following manner: for every*
(14) $\pi = <p_1, p_2, ..., p_m>$
(15) $f_r(\pi) = <p_m, p_1, p_2, ..., p_{m-1}>$
(16) $f_c(\pi) = <\xi(p_1), \xi(p_2), ..., \xi(p_m)>$
*where the function $\xi(a)$ is the defined in (13) function.*
*Let $A \in B_{m \times n}$ be a random $m \times n$ Boolean matrix and let*
(17) $B = \alpha(f_r(\alpha^{-1}(A)))$
(18) $C = \alpha(f_c(\alpha^{-1}(A)))$
*Then B is obtained from A by moving the last row to the first place, and C is obtained from A by moving the last column to the first place (respectively the first row or column becomes the second, the second becomes the third, respectively etc.).*

**Proof.** Let $\pi = <p_1, p_2, ..., p_m> = \alpha^{-1}(A) \in P_n^m$
Then the number $p_i$,
$0 \leq p_i \leq 2^n - 1$, $i=1, 2, ... , m$
will correspond to the $i$-th row of the matrix A. Then apparently the matrix
$B = \alpha(f_r(<p_1, p_2, ..., p_m>)) =$

$$= \alpha(< p_m, p_1, p_2, ..., p_{m-1} >)$$

is obtained from *A* by moving the last row in the place of the first one, and moving the remaining rows one row below.

Let $p_i \in P_n = \{0, 1, ..., 2^n – 1\}$, $i=1, 2, ...,m$. Then $d_i = p_i \% 2$ gives us the last digit of the binary representation of the number $p_i$. If $p_i$ is written in binary form with precisely *n* digits, optionally with insignificant zeros in the beginning, then by applying integer number division of $p_i$ to *2*, we practically remove the last digit $d_i$ and we move it to the first position, in case we multiply by $2^{n-1}$ and add it to $p_i/2$. This is how, by definition, the function $\xi(p_i)$ works. Hence, the $m \times n$ matrix

$$C = \alpha(f_c(< p_1, p_2, ..., p_m >)) =$$
$$= \alpha(< \xi(p_1), \xi(p_2), ..., \xi(p_m) >)$$

is obtained from the matrix *A* by moving the last column to the first position, and all the other columns are moved one column to the right.

From the definitions of the functions $f_r$ (15) and $f_c$ (16) it is easy to verify the validity of the following

**Proposition 2** *If by definition*

(19) $\qquad f_r^0(\pi) = f_c^0(\pi) = \pi$

(20) $\qquad f_r^k(\pi) = f_r(f_r^{k-1}(\pi))$

(21) $\qquad f_c^k(\pi) = f_c(f_c^{k-1}(\pi))$,

*where $\pi \in P_n^m$, k is a positive integer, then*

(22) $\qquad f_r^m(\pi) = \pi$

*and*

(23) $\qquad f_c^n(\pi) = \pi$

*for each $\pi \in P_n^m$.*

As a direct consequence of Theorems 1 and 2 and Proposition 2 and their constructive proofs, it follows that the following algorithm finds exactly one representative of each equivalence class with respect to the previously defined in problem 1 relation of equivalence $\rho$ and the cardinality of the factor set $B_{m \times n / \rho}$.

**Algorithm 2** We find at least one representative of each equivalence class $\rho$ and the cardinality of the factor set $B_{m \times n / \rho}$ when m and n are given.

*1. We declare the m-dimensional Boolean arrays W1 and W2 which we will be indexed by using the elements of the set $P_n^m$, i.e. W1[<$p_1,p_2,...,p_m$>] will correspond to the element <$p_1, p_2,..., p_m$> $\in P_n^m$ We proceed analogically with the array W2.*

*2. Initially we take all elements of W1 and W2 to be zeros. In W1 we will "remember" all elements chosen from $B_{m \times n}$ (one for each equivalence class) by changing W1[< $p_1, p_2,..., p_m$ >] to one if we have selected the element $\alpha(< p_1, p_2,...,p_m >)$ for a representative of the respective equivalence class. We will change the elements of W2 to 1 for each "crossing out" of an element from $B_{m \times n}$, i.e. for each $\pi'' \in P_n^m$, for which there exists $\pi' \in P_n^m$, such that W1[$\pi'$]=1 and $\alpha(\pi'')\rho\alpha(\pi')$, or in other words, by $\pi'$ and $\pi''$ two matrices of the same equivalence class are encoded as we have chosen $\alpha(\pi')$ for a representative of this equivalence class.*

*3. We declare the counter N, which we initialize by 0. In case of normal ending of the algorithm, N will be showing the cardinality of the factor set $B_{m \times n / \rho}$.*

*4. While a zero element exists in W2 do*
**Beginning** *of Cycle 1*
*5. We choose the minimum $\pi = < p_1, \pi_1,..., \pi_m > \in P_n^m$ according to the lexicographic order, for which W1 [$\pi$]=0.*
*6. We assign 1 to W1[$\pi$].*
*7. We increase N by one.*
*8. For i varying from 1 to m do*
**Beginning** *of Cycle 2*
*9. In place of the previous value of $\pi$ we assign a new value equal to $f_r^i(\pi)$.*
*10. For j varying from 1 to n do*
**Beginning** *of Cycle 3*
*11. In place of the previous value of $\pi$ we assign a new value equal to $f_r^i(\pi)$.*
*12. We assign one to W2[$\pi$].*
**End** *of Cycle 3*
**End** *of Cycle 2*
**End** *of Cycle 1*